\documentclass[aps,prb,twocolumn,superscriptaddress]{revtex4-1}
\usepackage{amsmath,amsthm,amssymb}
\usepackage[dvipdfmx]{graphicx}
\usepackage{amsmath,bm}
\usepackage{color,ulem}

\newcommand{\1}{\mbox{1}\hspace{-0.25em}\mbox{l}}
\newlength{\figwidth}
\newlength{\figlarge}
\begin{document}
\title{
Restoration of topological properties at finite temperatures in a heavy-fermion system
}

\author{Tsuneya Yoshida}
\affiliation{Condensed Matter Theory Laboratory, RIKEN, Wako, Saitama, 351-0198, Japan}
\author{Robert Peters}
\affiliation{Computational Condensed Matter Physics Laboratory, RIKEN, Wako, Saitama 351-0198, Japan}
\author{Norio Kawakami}
\affiliation{Department of Physics, Kyoto University, Kyoto 606-8502, Japan}

\date{\today}
\begin{abstract}
We study how topological phases evolve in the
  Kane-Mele-Kondo lattice at finite temperatures and
  obtain the topological Doniach phase diagram. 
In particular, we find an intriguing crossover behavior induced by
 the interplay between the topological structure and
electron correlations; the topological properties are restored by
temperature effects. This restoration can be observed
 in the behavior of the bulk as well as the edge.
In the bulk, we observe an increase of the spin-Hall
conductivity at finite temperatures, while it is zero in the low
temperature region. At the edge, we observe gapless edge modes emerging with increasing temperature.
\end{abstract}


\maketitle
\section{Introduction}
After the discovery of topological insulators, nontrivial properties of topological phases have been extensively studied. 
In these systems, various remarkable observations have been made 
because of the presence of gapless edge modes arising from topologically nontrivial bulk properties. 
\cite{Hasan10,Qi10} 
These modes induce 
the quantized Hall conductivity in integer quantum Hall systems and topological magnetoelectric effects in three-dimensional topological insulators. 
Furthermore, the edge mode in superconductors is the origin of exotic particles. 
Gapless edge modes in the one-dimensional
topological superconductor are composed of Majorana fermions whose
anti-particles are themselves, and 
their experimental realization has been addressed recently. \cite{Majorana_Mourik,Majorana_Rokhinson,Majorana_Das}

While topological insulators in free-fermion systems have been
extensively studied, the impact of correlations on these systems
becomes recently
one of the important issues in this field
\cite{AFvsTBI_DQMC_Hohenadler11,AFvsTBI_VCA_Yu11,AFvsTBI_CDMFTWu11,TBI_Mott_Yoshida,Raghu08,TM-Z_YZhang09}
because several $d$- and 
$f$-electron systems have been proposed as correlated
topological insulators.
\cite{Heusler_Chadov10,Heusler_Lin10,skutterudites_Yan12,TKI_SmB6_Takimoto,TKI_YbB12_Dai14,A2Ir2O7_Fiete15} 
In addition, the combination of topological structures and
electron correlations is expected to trigger off exotic
phenomena. An example is the topological Mott insulator
\cite{TMI_LBalents09}, where nontrivial properties of the
single-particle Green's function lead to gapless edge modes only in
the spin excitation spectrum. \cite{TMI_LBalents09,TMI_Yoshida14} 

The topological Kondo insulator \cite{TKI_Dzero10} is another example
of strongly correlated topological insulators. 
This phase
attracts much interest recently because there are various proposals
for topological insulators in heavy-fermion systems. It has been proposed that $\mathrm{SmB}_6$ is a realization of such a 
strong topological insulator, where
in-gap states have been observed via ARPES
measurements. \cite{SmB6_ARPESNeupane,SmB6_ARPESJiang,SmB6_ARPESXu}
$\mathrm{YbB}_{12}$ is proposed as a candidate material for 
topological
crystalline Kondo insulators. \cite{TKI_YbB12_Dai14} The topological phases in heavy-fermion systems have been
  extensively studied at zero temperature,
\cite{TKI_Dzero10,TKI_Dzero12,TKI_Yoshida13,TKL_Si13,TKL_Zhong13}
because the topological structure is well-defined in this limit.

An open question is, however, how finite temperatures
  affect topological Kondo insulators. This question is of particular interest
 because experiments for topological insulators are carried out at
 finite temperatures. 
For example, the Doniach phase diagram, which is well-known as a
  qualitative phase diagram for ordinary heavy-fermion systems, 
has not been established for topologically nontrivial systems.
Furthermore, at finite temperatures, we can expect peculiar behaviors due to the interplay between electron correlations and topologically nontrivial properties because characteristic temperature dependence has been observed even in the absence of topological properties.
Although several groups have studied topological Kondo insulators at finite temperatures,\cite{TKI_Tran12,TKI_Werner13,TKI_Werner14} temperature effects on topological phases have not been systematically explored.

In this paper, we elucidate how topologically nontrivial/trivial phases evolve at finite temperatures. 
Specifically, we study the Kane-Mele Kondo lattice at finite temperatures with the dynamical mean field theory (DMFT). 
The obtained Doniach phase diagram shows three
phases at zero temperature\cite{TKL_Si13,TKL_Zhong13}: an antiferromagnetic topological insulator, an antiferromagnetic trivial insulator, and a trivial Kondo insulator. 
Furthermore, we find a restoration of topologically
nontrivial properties at finite temperatures because of the interplay between the topologically nontrivial structure and
the electron correlations.
In order to study these phases, we analyze bulk and edge properties.
In the bulk, the spin-Hall conductivity which is almost zero around zero temperature increases with increasing temperature.
At the edge, the gapless edge modes emerge with increasing temperature. 
We elucidate that the interplay of the topological nature of the system and the Kondo effect is essential for the restoration of topological properties.

The rest of this paper is organized as follows. In the next section (Sec. \ref{sec: model_and_method}) we explain our model and method. 
The details of the results are discussed in Sec. \ref{sec: results}, which is followed by a short summary in Sec. \ref{sec: summary}.
\section{Model and method}\label{sec: model_and_method}
\begin{figure}[!h]
\begin{center}
\includegraphics[width=50mm,clip]{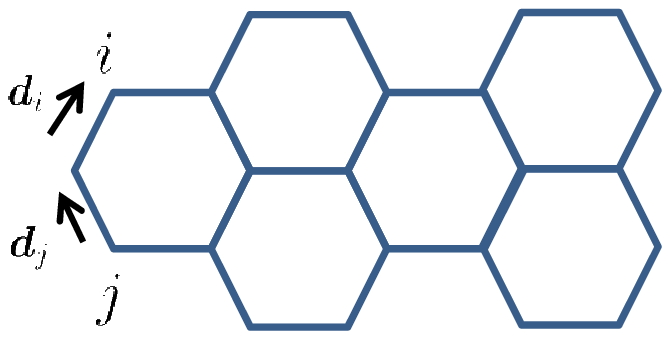}
\end{center}
\caption{(Color Online). 
Sketch of the lattice. The electron of the up-spin state hops from site $j$ to $i$ with $it_{so}\mathrm{sgn}(i,j)$, where $\mathrm{sgn}(i,j)=(\hat{\bm{d}}_i\times \hat{\bm{d}}_j)/|\hat{\bm{d}}_i\times \hat{\bm{d}}_j|$. 
}
\label{fig:model}
\end{figure}

In order to study topological properties in heavy-fermion systems, we analyze the Kane-Mele Kondo lattice. \cite{TKL_Si13,TKL_Zhong13} The Hamiltonian reads 
\begin{subequations}
\label{eq: HKMK}
\begin{eqnarray}
H_{KMK}&=& H_{KM}+J\sum_{i} {\bm s}_i \cdot {\bm S}_i,  \\
H_{KM} &=& -t\sum_{\langle i,j\rangle \sigma} c^\dagger_{i,\sigma}c_{j,\sigma} \nonumber \\
&&\;\;+it_{so}\sum_{\langle\langle i,j\rangle \rangle} \mathrm{sgn}(i,j)\mathrm{sgn}(\sigma) c^\dagger_{i,\sigma}c_{j,\sigma}, 
\end{eqnarray}
\end{subequations}
with ${\bm s}_i=\frac{1}{2}c^\dagger_{i,s}{\bm \sigma}_{s,s'}c_{i,s'}$ and $\mathrm{sgn}(\sigma)=1$ ($-1$) for $\sigma=\uparrow$ ($\downarrow$). 
Here $c^\dagger_{i,s}$ creates an electron with spin
$s=\uparrow,\downarrow$ at site $i$. ${\bm S}_{i}$ denotes a localized
moment of spin $S=1/2$ 
at site $i$. 
$\mathrm{sgn}(i,j)=\hat{\bm{d}}_i\times \hat{\bm{d}}_j/|\hat{\bm{d}}_i\times \hat{\bm{d}}_j|$, where $\bm{d}_i$ and $\bm{d}_j$ are vectors connecting site $i$ and $j$ (see Fig.~\ref{fig:model}). 
Namely, it takes 
$1$ ($-1$) when the electron hops clockwise (counter clockwise), respectively. 
The second term in Eq.~{(\ref{eq: HKMK}a)} denotes the antiferromagnetic ($J>0$) coupling between the localized spin and the itinerant electron at the same site.
This model is 
 defined on the 
honeycomb lattice and has two sublattices, A- and B-site. 
The next nearest hopping $t_{so}$, whose characteristic form originates from the spin-orbit interaction, results in the bulk band gap for $J=0$, thereby stabilizing a topological insulating phase for the conduction electrons.

In order to analyze the system, we employ DMFT, which enables us to treat local correlation exactly. \cite{DMFT_Georges,DMFT_Metzner,DMFT_Muller-Hartmann}
With this approach, we map the system onto an effective impurity
problem and solve it self-consistently. The self-consistency equation is given by
\begin{eqnarray}
 \hat{g}^{-1}_{\sigma}(\omega)&=& [\sum_{{\bm k}}\frac{1}{(\omega+i\delta) \1-\hat{h}_{\sigma}({\bm k})-\hat{\Sigma}^{R}_{\sigma}(\omega)} ]^{-1} +\hat{\Sigma}^{R}_{\sigma}(\omega), \nonumber \\
\end{eqnarray}
where $\hat{h}_{\sigma}({\bf k})$ is the two-dimensional matrix obtained by the Fourier transform of the hopping matrix. 
The self-energy $\hat{\Sigma}^R_{\sigma}(\omega)$ of the retarded Green's function for the lattice system is a diagonal matrix and can be computed from the Green's function $\hat{g}_{\sigma}(\omega)$ of the effective impurity model. 
In this study, we employ the numerical renormalization group (NRG)
method for the computation of the self-energy,
\cite{NRG_Wilson,NRG_Peters,NRG_Weichselbaum,NRG_Bulla} which is
appropriate for calculations in the low temperature region.

For $J=0$, our system is reduced to the ordinary Kane-Mele model. In this case, the topological structure of the ground state is characterized by the spin Chern number $N_{\mathrm{sCh}}$, which is proportional to the spin-Hall conductivity. 
In terms of the Green's function, this topological invariant is defined as follows: \cite{Volovik03,Gurarie11}
\begin{eqnarray}
N_{\mathrm{sCh}} &=& \frac{\epsilon^{\mu \nu \rho}}{48\pi^2} \int d\omega d\bm{k}\; \mathrm{sgn}(\sigma)\mathrm{tr}[ 
\hat{G}^{-1}_{\sigma}(k) \partial_{\mu}  \hat{G}_{\sigma}(k) \nonumber \\
&& \quad \quad \hat{G}^{-1}_{\sigma}(k) \partial_{\nu}  \hat{G}_{\sigma}(k)
\hat{G}^{-1}_{\sigma}(k) \partial_{\rho} \hat{G}_{\sigma}(k)
], \label{eq: def_sCh}
\end{eqnarray}
where $\epsilon^{\mu\nu\rho}$ ($\mu,\nu,\rho$ $=0,1,2$) is the anti-symmetric tensor taking $\epsilon^{012}=1$, $\bm{k}:=(k_x,k_y)$, $\bm{\partial}:=(\partial_\omega,\partial_{k_x},\partial_{k_y})$. 
Here, $\hat{G}_{\sigma}(k):=\hat{G}_{\sigma}(i\omega,\bm{k})$ is the two-dimensional matrix in the pseudospin (sublattice) space and defined along the imaginary axis.
This quantity characterizes the map from $(\omega,\bm{k})$ to $GL(2,\mathbb{C})$, and is well-defined even in correlated systems. \cite{Gurarie11}
For a continuous change of the ground state wave function, the topological structure changes only when the Green's function does not belong to $GL(2,\mathbb{Z})$.
Namely, if the topological invariant changes with a continuous transition, the Green's function satisfies
\begin{subequations}
\label{eq: green_singlar}
\begin{eqnarray}
\mathrm{det}\hat{G}^{-1}_{\sigma}&=&0,
\end{eqnarray}
or 
\begin{eqnarray}
\mathrm{det}\hat{G}_{\sigma}&=&0, 
\end{eqnarray}
\end{subequations}
at a certain point of the space $(\omega,\bm{k})$. 
Physically Eq.~{(\ref{eq: green_singlar}a)} can be satisfied by
gap-closing in the single-particle excitation spectrum. The
topological transition satisfying Eq.~{{(\ref{eq: green_singlar}b)}}
is possible only for correlated systems. Without correlations,
it is equivalent to a diverging band width. In the presence of electron
correlations, Eq.~{{(\ref{eq: green_singlar}b)}} can be satisfied by a 
divergence of the self-energy. 

Throughout this paper, we set the model parameters to $t=1$ and
$t_{so}=0.1$, and take the distance of neighboring sites as the unit of length. 
For the Green's function along the imaginary axis, we calculate it for
Matsubara frequencies $i\omega_n=(2n+1)\pi T$ with $n\in \mathbb{Z}$
and $T=0.0001t$. The size of the bulk-gap $\Delta$ is
$\Delta\sim0.01t$, and thus much larger than the temperature, $T=0.0001$.

\section{Results}\label{sec: results}
\subsection{Doniach phase diagram}
\begin{figure}[!h]
\begin{center}
\includegraphics[width=\hsize,clip]{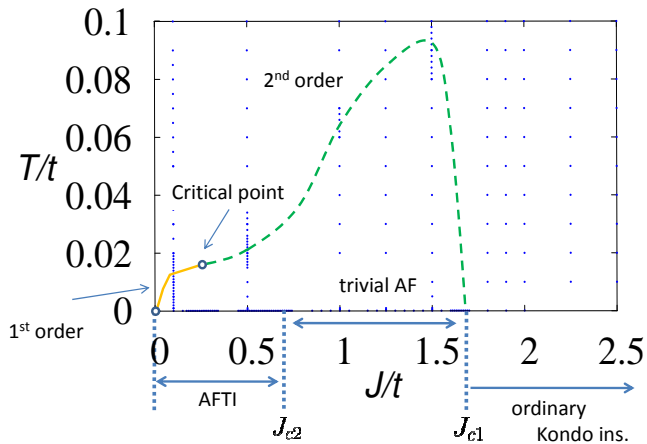}
\end{center}
\caption{(Color Online). 
Phase diagram of temperature ($T$) vs. the antiferromagnetic coupling ($J$). 
Because of the presence of the bulk-gap, a first order transition (solid orange line) is observed in the weak-coupling region while it changes to a second order (dashed green line) as increasing the coupling.
In this figure, one can observe the topological phase transition. For $J< 0.683t$ a topological antiferromagnetic phase (AFTI) emerges. This phase changes into a trivial antifferomagnetic phase and an ordinary Kondo insulator as the interaction is increased.
}
\label{fig:phasediagram}
\end{figure}
Let us start with explaining the Doniach phase diagram. 
The calculated results are summarized in Fig.~\ref{fig:phasediagram}.
For weak exchange interaction $J$, we can find an antiferromagnetic
topological phase, which is induced by the RKKY interaction. 
This phase changes to the trivial antiferromagnetic phase with
increasing $J$. With increasing $J$, the Kondo effect
  becomes dominant and the topologically trivial Kondo insulator (i.e., the ordinary Kondo insulator)
  appears. 
 Increasing temperature destroys the antiferromagnetic order with a first (second) order transition for weak (strong) interaction strength $J$. 
 In the following, we discuss the details of the phase diagram.

\begin{figure}[!h]
\begin{center}
\includegraphics[width=\hsize,clip]{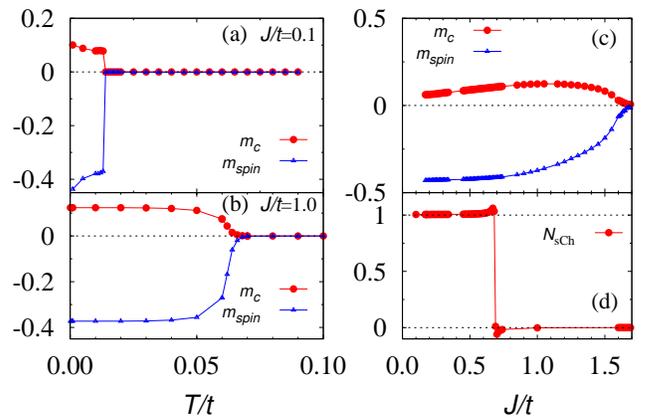}
\end{center}
\caption{(Color Online). 
The spin Chern number and magnetic moments.
In panels (a) and (b), temperature dependence of the antiferromagntic moments is plotted for $J=t$ and $J=0.1t$, respectively.
$m_c$ and $m_{spin}$ denote the magnetic moment of the itinerant electron and localized spin. 
(c): the antiferromagntic moment as functions of the interaction strength.
(d): the spin Chern number as a function of the interaction strength $J$.
}
\label{fig:mag}
\end{figure}

First, we focus on the 
properties of the magnetic phases in Fig.~\ref{fig:phasediagram}. 
If the interaction strength $J$ is weak, the RKKY interaction is dominant, and we can find the magnetic phase. 
Besides, if $J$ is sufficiently weak, a first order transition is
observed as a function of temperature, the origin of which can be
  found in the bulk-gap. 
The temperature dependence of the magnetization is plotted in Fig.~{\ref{fig:mag}(a)} and (b) for $J=0.1t$ and $J=t$, respectively. For $J=0.1t$ the magnetization shows a jump at $T=0.013t$ [see Fig.~{\ref{fig:mag}(a)}]. 
Here we have defined the antiferromagnetic moment for the conduction electron and localized spin at A-site as $m_c=\langle n_{i,\uparrow}-n_{i,\downarrow}\rangle/2$ and $m_{spin}=\langle S^z_i \rangle$ with $i\in$[A-site], respectively. 
With increasing $J$, this phase transition becomes second order; the
Kondo temperature becomes energetically larger than the
band gap 
and a continuous transition is observed for $J=t$ [see Fig.~{\ref{fig:mag}(b)}]. 
Correspondingly the quantum phase transition at $J_{c1}\sim 1.61t$ is also continuous; the magnetization around zero temperature ($T=0.0001t$) approaches zero continuously [see Fig.~{\ref{fig:mag}(c)}].

Next, we discuss the topological structure of the phase diagram. 
The antiferromagnetic phase can possess topologically nontrivial properties. \cite{AFTI_Yoshida13,TKL_Zhong13}
We can use a simplified formula \cite{Wang12} for the spin Chern
number, because the Green's function is not singular in this magnetic
phase as we will show 
in the next subsection and the Appendix;
\begin{subequations}
\label{eq: simplified_sCh}
\begin{eqnarray}
N_{\mathrm{sCh}} &=& \sum_{\sigma} \frac{\mathrm{sgn}(\sigma)}{8\pi} \int d\bm{k}\hat{\bm{d}}_{\sigma}(\bm{k}) \cdot \partial_{k_x}\hat{\bm{d}}_{\sigma}(\bm{k}) \times  \partial_{k_y}\hat{\bm{d}}_{\sigma}(\bm{k}), \nonumber \\
\end{eqnarray}
with
\begin{eqnarray}
\hat{h}'_{\sigma}(\bm{k}) &:=& \hat{h}_{\sigma}(\bm{k})+\hat{\Sigma}_{\sigma}(i\omega=0), \nonumber \\
                          &=& d_{\sigma,0}(\bm{k}) \1  +\sum_{i}d_{\sigma,i}(\bm{k})\hat{\tau}_i,
\end{eqnarray}
\end{subequations}
and $\hat{d}_i=d_i/|d|$.
Here $\1$ and $\hat{\tau}_i$ ($i=x,y,z$) denote the two-dimensional identity matrix and the Pauli matrix for the sublattice space. 
We have confirmed that the results are consistent with those obtained from the definition [Eq.~{(\ref{eq: def_sCh})}]. The results are plotted in Fig.~{\ref{fig:mag}(d)}.

\begin{figure}[!h]
\begin{center}
\includegraphics[width=\hsize ,clip]{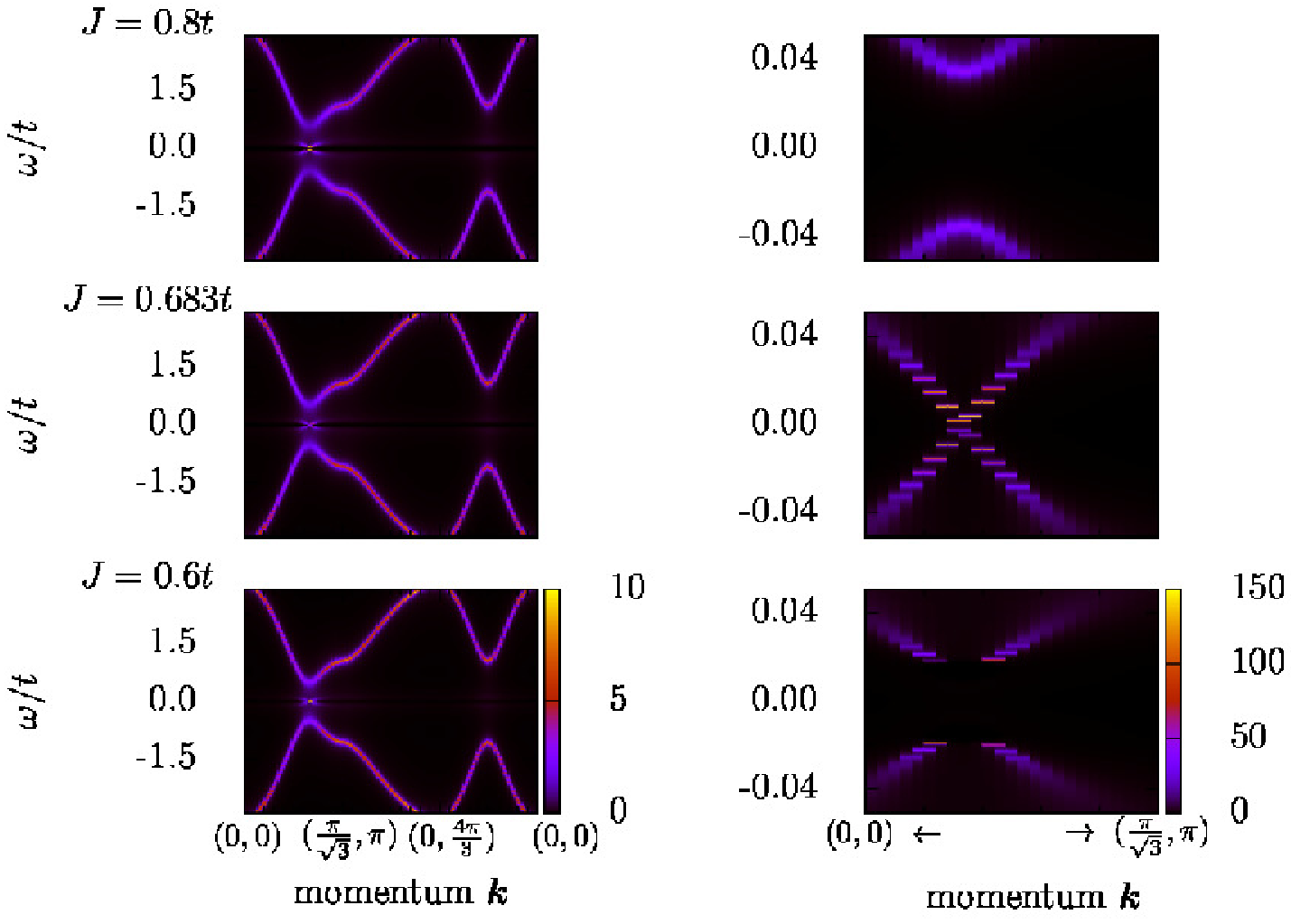}
\end{center}
\caption{(Color Online). 
Momentum-resolved spectral function $A(\omega,\bm{k})=-\mathrm{Im}\sum_{\sigma}\mathrm{tr}\hat{G}^R_{\sigma}(\omega,\bm{k})/\pi$ for several values of the interaction strength $J$. 
In this figure the momentum $\bm{k}$ sweeps the high symmetry points in the Brillouin zone; $(k_x,k_y)=(0,0)$ $\to$ $(\pi/\sqrt{3},\pi)$ $\to$ $(0,4\pi/3)$ $\to$ $(0,0)$. 
Here we have taken the distance of neighboring sites as the length unit. 
The data for $J=0.6t$, $0.683t$, and $0.8t$ are plotted from bottom to top. Right panels are the magnified version around $\omega=0$.
}
\label{fig:Ak}
\end{figure}

As seen in this figure, we find that the system shows the
topologically nontrivial properties for $J< 0.683t$, while a further increase of the interaction strength induces a topological transition. At this continuous topological transition point, gap-closing in the density of states is observed; the Green's function becomes singular, satisfying Eq.~{(\ref{eq: green_singlar}a)} at $i\omega=0$.
The gap-closing is shown 
in Fig.~\ref{fig:Ak}.
In this figure, the momentum-resolved spectral function
$A(\omega,\bm{k})=-\mathrm{Im}\sum_{\sigma}\mathrm{tr}\hat{G}^R_{\sigma}(\omega,{\bm
  k})/\pi$ is plotted for $J=0.6t$, $ 0.683t$, and $ 0.8t$ from bottom
to top. The right panels are magnified versions of them. 
As it is observed in the ordinary Kondo lattice model, a renormalized
band appears around $\omega=0$ because of the correlations between the conduction electrons and the 
localized spins. When 
approaching the topological phase transition point from the nontrivial
phase ($J=0.6t$), we can see that the bulk-gap becomes firstly narrower
($J_{c2}=0.683t$), and then widens again 
with further increasing of $J$ ($J=0.8t$).
A difference in the origin of the gap formation between these two phases gives rise to this gap-closing. 
For $J<J_{c2}$ the bulk gap arises from the hopping $t_{so}$, while
for $J>J_{c2}$, the antiferromagnetic order induces the gap.
Here, $J_{c2}$ denotes the critical value for the topological phase transition.
Putting the above results together, we end up with the Doniach phase diagram shown in Fig.~{\ref{fig:phasediagram}}.

We finish this subsection with a comment on the topological structure in the Kondo lattice. 
As discussed in the next subsection in detail, there is no topological phase if
the system is paramagnetic because zeros of the single-particle Green's function (i.e., pole of the self-energy) appear. 
On the other hand, in Refs.~\onlinecite{TKL_Si13} and~\onlinecite{TKL_Zhong13}, the nontrivial phase persists even in the paramagnetic phase. 
This apparent contradiction arises from the difference of the used methods; in Refs.~\onlinecite{TKL_Si13} and~\onlinecite{TKL_Zhong13} the localized spin is treated with the slave-boson approach and 
is separated into a boson and a fermion which
participate in the topological structure. 
The authors study the resultant model at the mean-field level, obtaining the paramagnetic topological phase.
However, 
if the local correlation is treated exactly, the topological structure in
the paramagnetic phase is destroyed due to zeros of the Green's
function in the limit of the Kondo lattice.
On the other hand, for small $J$, the RKKY interaction induces the antiferromagnetic order and removes the zeros of the Green's function. 
Thus, in the Kane-Mele Kondo lattice model, the nontrivial state is stabilized only in the antiferromagnetic phase.

\subsection{Restoration of topological properties at finite temperatures}
In this section we will demonstrate that in heavy-fermion systems with topological
structure, the interplay between electron correlations
and topological properties leads to an intriguing crossover behavior;
topological properties are restored at finite temperatures in the
region where the topological structure of the ground state is 
destroyed.
This behavior is observed via analysis of 
the bulk and the edge.
Calculations for the bulk systems reveal that the spin-Hall
conductivity increases even if the topological structure is absent 
at zero temperature.
Calculations under ribbon geometry [the open (periodic) boundary condition for $x$- ($y$-) direction] reveal that gapless edge modes appear due to temperature effects.
In the following, we will confirm these intriguing properties in two steps.
\begin{figure}[!h]
\begin{center}
\includegraphics[width=65mm,clip]{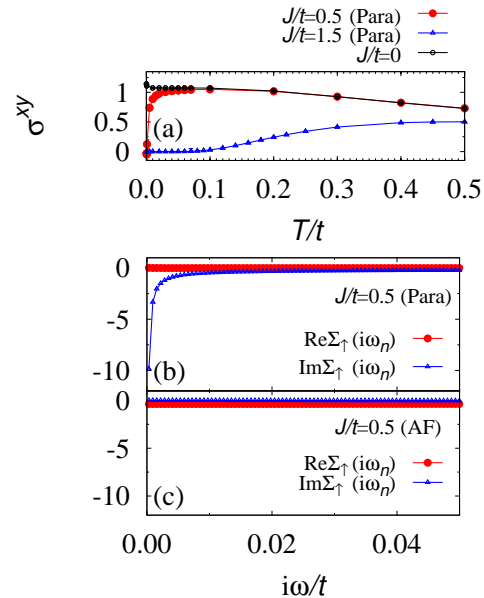}
\end{center}
\caption{(Color Online). 
(a): temperature dependence of the spin-Hall conductivity. (b) ((c)): self-energy of up-spin state in the paramagnetic (antiferromagnetic) phase, respectively. Here, we plot them along the imaginary axis. In the antiferromagnetic phase, we plot the self-energy at the A-site where the electron state for up-spin is majority.
}
\label{fig:cond_T}
\end{figure}

In order to see the essence of the restoration of topological properties, let us first restrict ourselves to paramagnetic solutions.
Temperature dependence of the conductivity is plotted in Fig.~{\ref{fig:cond_T}(a)}.
In the low temperature region, the spin-Hall conductivity is zero since the topological invariant is no longer well-defined due to the Kondo effect; the singlet formation between electrons and localized spins leads to zeros of the Green's function (i.e., divergence of the self-energy) [see Fig.~{\ref{fig:cond_T}(b)}].\cite{footnote}
With increasing temperature, the Kondo effect is suppressed, and the conductivity increases.
For $T\gtrsim 0.03t$, the conductivity approaches the values of $J=0$ which are almost quantized.
The increase of the spin-Hall conductivity is also observed for $J=1.5t$, even though increasing the coupling strength $J$ suppresses the conductivity at finite temperatures due to enhancement of the Kondo effect.

\begin{figure}[!h]
\begin{center}
\includegraphics[width=90mm]{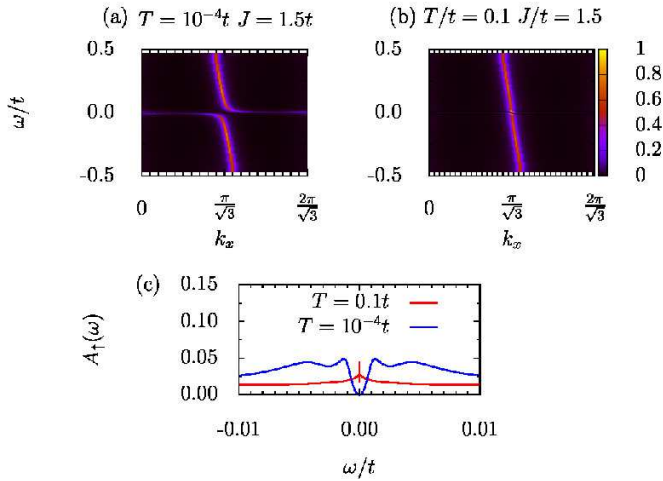}
\end{center}
\caption{
Momentum-resolved spectral function for up-spin state $A_{\uparrow}(\omega,\bm{k})=-\mathrm{Im}\;\mathrm{tr}\hat{G}^R_{\uparrow}(\omega,\bm{k})/\pi$ at the edge site $x=0$ for $T=0.1t$ and $T=0.0001t$.
In the spectral weight for down-spin state, we can find edge modes propagating in opposite direction.
The R-DMFT is performed under the open (periodic) boundary condition for $x$- ($y$-) direction. We have zig-zag edges at $x=0$ and $L-1$. 
For $T=0.0001t$ edge modes are gapped due to the Kondo effect, while the gap disappears with increasing temperature.
For $T=0.1t$, an extremely tiny gap is observed. 
We conclude that this is due to the finite size effect because the gap width should be the same as that for $T=0.0001$ if the Kondo effect induces it.
}
\label{fig:Ak_lay_paper}
\end{figure}
This increase of the conductivity is interpreted as a restoration of gapless edge modes. 
Our real-space dynamical mean-field theory (R-DMFT) calculations using
the ribbon geometry reveal how finite temperatures affect the edge modes. 
In Fig.~{\ref{fig:Ak_lay_paper}} (a) and (c), we can see that edge modes are destroyed due to the Kondo effect for $T=0.0001t$, which is consistent with the bulk behavior.
With increasing temperature, the gap in the local density of states at
the edge site disappears [see
  Fig.~\ref{fig:Ak_lay_paper}(c)]. Correspondingly, we can observe a restoration of edge modes in
Fig.~\ref{fig:Ak_lay_paper}(b), which leads to an increase of the spin-Hall conductivity.
\begin{figure}[!h]
\begin{center}
\includegraphics[width=70mm,clip]{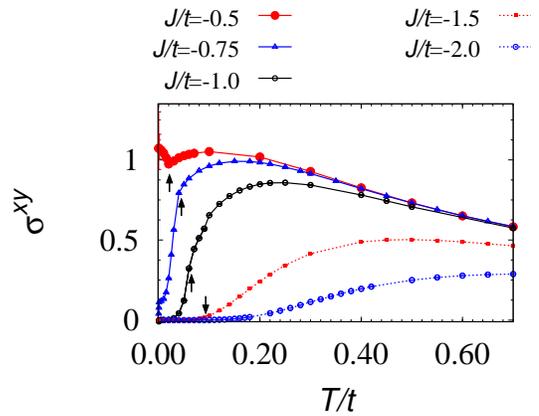}
\end{center}
\caption{(Color Online). 
Temperature dependence of the spin-Hall conductivity for several values of coupling strength $J$.
The error bars in the low temperature region arise from extrapolation to the $\omega\to 0$ limit.
At high temperatures these error bars are smaller than symbols.
Arrows denote antiferromagnetic transition points.
}
\label{fig:cond_TvarJ}
\end{figure}

We have so far demonstrated 
the characteristic crossover behavior at finite temperatures in
heavy-fermion systems. 
The origin of this crossover
  is the competition of the Kondo effect and the topological properties.
In the low temperature region, the Kondo effect governs 
  the low energy properties and destroys the topological structure.
On the other hand, topological properties (i.e., the conductivity and
edge states) are restored if the temperature is higher than the Kondo
temperature but smaller than the energy scale of the
band gap of the 
non-interacting topological insulator.

An important point to be noticed here is that the restoration of topological properties is also observed in the antiferromagnetic phase. Namely, the increase of the spin-Hall conductivity is also observed in the antiferromagnetic phase. 
In Fig.~\ref{fig:cond_TvarJ}, temperature dependence of the conductivity in the antiferromagnetic phase is plotted.
For $J=0.75t$, the ground state is topologically trivial. 
Correspondingly, the spin-Hall conductivity vanishes at zero temperature, but increases with increasing temperature.
We note that the ground state is an 
antiferromagnetic topological phase for $J<0.683$. 
As seen in Fig.~\ref{fig:cond_T}(c), the magnetic order removes the
pole of the self-energy, and the ground state possesses nontrivial
properties (properties of the self-energy in the trivial
antiferromagnetic phase are 
discussed in Appendix~\ref{sec: app_self}).
This topological phase leads to a dip structure in the temperature dependence of the conductivity for $J=0.5t$; 
with decreasing temperature the Kondo effect firstly decreases the conductivity for $0.021t<T\lesssim 0.03t$, but with entering the antiferromagnetic phase ($T<0.021t$), the conductivity increases again because the ground state is the antiferromagnetic topological insulator.

\section{Conclusion}\label{sec: summary}
In this paper, we have analyzed the Kane-Mele Kondo lattice model at finite temperatures and obtained the Doniach phase diagram under the topologically nontrivial condition (Fig.~{\ref{fig:phasediagram}}). 

In particular, we have observed an 
intriguing crossover behavior due
to the interplay between electron correlations and topological
properties. In the low temperature region, the Kondo effect destroys
the topological structure, while the topological properties are
restored if the temperature is higher than the Kondo temperature but
smaller than the band gap of the topological insulator in the
non-interacting case. The restoration of topological properties can be observed in the bulk and the edge.
The spin-Hall
conductivity rapidly increases even if it is almost zero in the low
temperature region. Edge modes, destroyed by the Kondo effect in the
low temperature region, appear with increasing temperatures. The
crossover behavior is observed in both of the paramagnetic and the
antiferromagnetic phases. 


\section{Acknowledgments}\label{sec: acknowledgements}
This work is partly supported by a Grand-in-Aid for Scientific Research on Innovative Areas (KAKENHI Grant No. 15H05855) and also KAKENHI (No. 25400366). 
The numerical calculations were performed at the ISSP in the University of Tokyo and on the SR16000 at YITP in Kyoto University.

\appendix
\section{SELF-ENERGY IN THE ANTIFERROMAGNETIC PHASE}\label{sec: app_self}
In this appendix, we discuss why the topological antiferromagnetic phase is not stabilized around the quantum critical point ($J_{c1}\sim1.61t$) in spite of the small magnetization.
By using the Hartree approximation, the self-energy is estimated as
$\Sigma_{\uparrow}\sim Jm_c$ in the antiferromagnetic
phase. Furthermore, 
around the quantum critical point $J_{c1}\sim1.61t$, the magnetic moment approaches zero.
With these facts, one might expect that a topological phase can be stabilized.
This is, however, not the case. Although the self-energy is
non-singular in the antiferromagnetic phase, a huge
  real-part of the self-energy, $-\mathrm{Re}\Sigma_\uparrow$, drives
the systems into the trivial phase around the critical
point. In the following we will discuss 
details.

\begin{figure}[!h]
\begin{center}
\includegraphics[width=60mm]{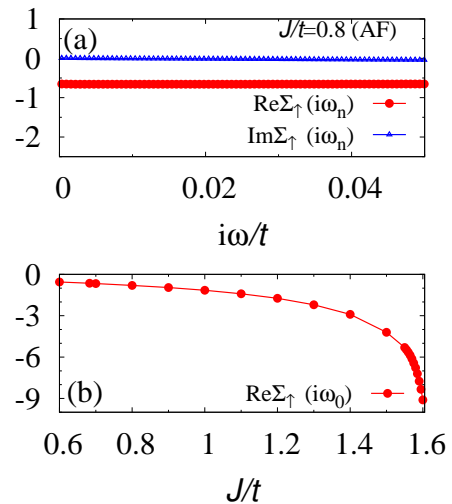}
\end{center}
\caption{(a) self-energy at $J=0.8t$. (b) the real part of the self-energy at $i\omega_0$. }
\label{fig:self_trivialAF}
\end{figure}

First, in the 
absence of a singularity of the Green's function the topological structure is well-defined. 
In Fig.~\ref{fig:self_trivialAF}(a), the self-energy for $J=0.8t$ is
plotted. 
In Figs.~\ref{fig:cond_T}(c) and\ \ref{fig:self_trivialAF}(a) the
self-energy shows no divergence in both the 
topologically nontrivial and trivial phases. 
With increasing $J$, $-\mathrm{Re}\Sigma_\uparrow(i\omega_0)$ becomes larger, which induces the gap-closing in the antiferromagnetic phase at $J_{c2}\sim 0.683t$.

Second we show that the real part of the self-energy,
$-\mathrm{Re}\Sigma_\uparrow(i\omega_0)$, rapidly increases when 
approaching the quantum critical point $J_{c1}$, see Fig.~\ref{fig:self_trivialAF} (b).
This figure indicates that $-\mathrm{Re}\Sigma_\uparrow (i\omega_0)$ increases rapidly with approaching the critical point even though the magnetic moment becomes small.
The large value of $-\mathrm{Re}\Sigma_\uparrow(i\omega_0)$ induces 
a 
bulk-gap and destroys the gap induced by 
the next-nearest-neighbor hopping
at $J_{c2}=0.683t$.
Hence, we cannot find a topological phase around the quantum critical point.

Therefore, we conclude that the system is in a topologically trivial phase around $J_{c1}$ because of a huge value of $-\mathrm{Re}\Sigma(i\omega_0)$.



\begin{thebibliography}{99}
\bibitem{footnote}
The ground state for $J>0$ is never obtained by applying perturbative expansion in the coupling $J$ to the ground state of $J=0$. 
As a consequence, the self-energy shows a divergence (i.e. the Green's function becomes zero).
\end{thebibliography}
\end{document}